\documentclass[11pt]{article}
\usepackage {amsfonts}
\usepackage {graphicx}
\usepackage {epsfig}
\usepackage{amssymb}
\usepackage{amsmath}
\usepackage {longtable}
\usepackage[english]{babel}
\begin{document}

\title{About influence of the deuteron electric and magnetic plarizabities on measurement of the
deuteron EDM in a storage ring}

\author{V.G. Baryshevsky \\ Research Institute for Nuclear Problems, Belarusian State
University,\\ 11 Bobruyskaya Str., Minsk 220050, Belarus,
\\ e-mail: bar@inp.minsk.by}

\maketitle

\begin{center}
\begin{abstract}
In the present paper influence of tensor electric and
 magnetic polarizabilities on spin evolution in the resonance deuteron EDM
 experiment  is considered in details.

 It is shown that besides EDM the electric and magnetic polarizabilities also contribute
 to the vertical spin component $P_3$. Moreover, the electric polarizability contributes
 to the $P_3$ component even when the deuteron EDM is supposed to be zero and thereby
 the electric polarizability can imitate the EDM contribution.
 It is shown that unlike the vertical component of the spin $P_3$
 the component $P_{33}$ of polarization tensor does not contain
 contribution from the electric polarizability, whereas contribution from the magnetic
 polarizability reveals only when the deuteron EDM differs from zero.

Moreover, it is also shown that when the angle $\vartheta$ between
the spin direction and the vertical axis meets the condition $\sin
\vartheta=\sqrt{\frac{2}{3}}$ ($\cos
\vartheta=\sqrt{\frac{1}{3}}$), the initial value of $P_{33}$
appears $P_{33}(0)=0$.
As a result, EDM contribution to the measured signal linearly
growth in time starting from zero that is important for
measurements.

Therefore, measurement of the $P_{33}$ component of deuteron
tensor polarization seems to be of particular interest, especially
because  the nonzero component $P_{33}$ appearance on its own
indicates the EDM presence (in contrast to the $P_3$ component,
which appearance can be aroused by the tensor electric
polarizability, rather than EDM).
\end{abstract}
\end{center}

\section{INTRODUCTION}
At present time the possibility to measure the electric dipole
moment of a deuteron moving in a storage ring is actively
discussed \cite{orlov,6}.
According to \cite{orlov,6} two types of experiments are discussing now:
a). deuteron energy is chosen to zeroize $(g-2)$ precession ($g$
is the gyromagnetic ratio)
b). deuteron beam velocity is modulated with the frequency
$\Omega_f$ close to the $(g-2)$ precession frequency $\Omega$, as
a result, this makes possible to observe the EDM signal as growth
in time of the vertical spin component \cite{orlov,6}.

These methods can provide for EDM measurement the sensitivity
$\sim 10^{-29}~e \cdot cm$.
Theoretical description of the experiment \cite{orlov,6} is being
done on the base of the Bargman-Myshel-Telegdy equation.
But as it is shown in \cite{nastya}-\cite{birefringence} the
Bargman-Myshel-Telegdy equation can not describes deuteron spin
behavior in such experiments.
It turns out that the BMT equation for a deuteron should be supplemented with
several additions, which describe interaction of deuteron electric
and magnetic polarizabilities with the electric field in the
storage ring and deuteron birefringence in matter.

Owing to the above in the experiments planned for the deuteron EDM
search the contributions aroused by { {the tensor electric and
magnetic polarizabilities of deuteron as well as the
spin-dependent amplitude of forward scattering by the nuclei of a
solid or gas target}} will be also measured.
These contributions could distort the EDM signal and even bring to
wrong conclusions about EDM observation.
They are the systematic errors for the EDM search, which should be
 eliminated.

 At the same time the above effects being measured in the experiments for
 EDM search could be even used for more reliable limits assignment
 for deuteron EDM.

 In the present paper influence of tensor electric and
 magnetic polarizabilities on spin evolution in the resonance deuteron EDM
 experiment  is considered in details.

 It is shown that besides EDM the electric and magnetic polarizabilities also contribute
 to the vertical spin component $P_3$. Moreover, the electric polarizability contributes
 to the $P_3$ component even when the deuteron EDM is supposed to be zero and thereby
 the electric polarizability can imitate the EDM contribution.
 It is shown that unlike the vertical component of the spin $P_3$
 the component $P_{33}$ of polarization tensor does not contain
 contribution from the electric polarizability, whereas contribution from the magnetic
 polarizability reveals only when the deuteron EDM differs from zero.

Moreover, it is also shown that when the angle $\vartheta$ between the spin
direction and the vertical axis meets the condition $\sin
\vartheta=\sqrt{\frac{2}{3}}$ ($\cos
\vartheta=\sqrt{\frac{1}{3}}$), the initial value of $P_{33}$
appears $P_{33}(0)=0$.
As a result, EDM contribution to the measured signal linearly
growth in time starting from zero that is important for
measurements.
Therefore, measurement of the $P_{33}$ component of deuteron
tensor polarization seems to be of particular interest, especially
because  the nonzero component $P_{33}$ appearance on its own
indicates the EDM presence (in contrast to the $P_3$ component,
which appearance can be aroused by the tensor electric
polarizability, rather than EDM).

\section{Interactions contributing to the spin motion of a particle in a storage ring}

As it is shown in \cite{nastya}-\cite{birefringence} considering evolution of the spin of a
particle in a storage ring, when measure EDM, one should take
into account several interactions:

1. interactions of the magnetic and electric dipole moments with
an electromagnetic field;

2. interaction of the particle with the  electric field due to the
tensor electric polarizability;

3. interaction of the particle with the magnetic field due to the
tensor magnetic polarizability;

4. interaction of the particle with the pseudoelectric nuclear
field of matter.

The equation for the particle spin wavefunction considering all
these interactions is as follows:
\begin{equation}
i\hbar\frac{\partial\Psi(t)}{\partial
t}=\left(\hat{H}_{0}+\hat{V}_{EDM}+\hat{V}_{\vec{E}}+\hat{V}_{\vec{B}}+\hat{V}_E^{nucl}\right)\Psi(t) \label{1}
\end{equation}
where $\Psi(t)$ is the particle spin wavefunction,

{$\hat{H}_{0}$ is the Hamiltonian describing the spin behavior
caused by interaction of the magnetic moment with the
electromagnetic field (equation (\ref{1}) with the only
$\hat{H}_{0}$  summand converts to the Bargman-Myshel-Telegdy
equation),}

$\hat{V}_{EDM}$ describes interaction of the particle EDM $d$ with the
electric field,
\begin{eqnarray}
\hat{V}_{EDM} & = &
-d\left(\vec{\beta}\times\vec{B}+\vec{E}\right)\vec{S},
\label{VEDM}
\end{eqnarray}
$\vec{\beta}=\frac{\vec{v}}{c}$, $\vec{v}$ is the particle velocity, $c$ is the speed of light.

$\hat{V}_{\vec{E}}$ describes interaction of the particle with the
electric field due to the tensor electric polarizability:
\begin{equation}
\hat{V}_{\vec{E}}=-\frac{1}{2}\hat{\alpha}_{ik}(E_{eff})_{i}(E_{eff})_{k},
 \label{VE}
\end{equation}
where $\hat{\alpha}_{ik}$ is the electric polarizability tensor of
the particle  , $\vec{E}_{eff}=(\vec{E}+\vec{\beta} \times
\vec{B})$ is the effective electric field; the expression
(\ref{VE}) can be rewritten as follows:
\begin{eqnarray}
\hat{V}_{\vec{E}} =
\alpha_{S}E^{2}_{eff}-\alpha_{T}E^{2}_{eff}\left(\vec{S}\vec{n}_{E}\right)^{2},~
\vec{n}_{E}  =
\frac{\vec{E}+\vec{\beta}\times\vec{B}}{|\vec{E}+\vec{\beta}\times\vec{B}|}
 \label{VE1}
\end{eqnarray}
where $\alpha_{S}$ is the scalar electric polarizability and
$\alpha_{T}$ is the tensor electric polarizability of the
particle.

A deuteron also has the magnetic
polarizability which is described by the magnetic polarizability
tensor $\hat{\beta}_{ik}$. Interaction of the particle with the
magnetic field due to the tensor magnetic polarizability is as
follows:
\begin{equation}
\hat{V}_{\vec{B}}=-\frac{1}{2}\hat{\beta}_{ik}(B_{eff})_{i}(B_{eff})_{k},
 \label{VB}
\end{equation}
where $(B_{eff})_{i}$ are the components of the effective magnetic
field $\vec{B}_{eff}=(\vec{B}-\vec{\beta} \times \vec{E})$;
$\hat{V}_{\vec{B}}$ (\ref{VB}) could be expressed as:
\begin{equation}
\hat{V}_{\vec{B}}=\beta_{S}B_{eff}^{2}-\beta_{T}B_{eff}^{2}\left(\vec{S}\vec{n}_{B}\right)^{2},~
\vec{n}_B=\frac{\vec{B}-\vec{\beta} \times
\vec{E}}{|\vec{B}-\vec{\beta} \times \vec{E}|}.
 \label{VB1}
\end{equation}
where $\beta_{S}$ is the scalar magnetic polarizability and
$\beta_{T}$ is the tensor magnetic polarizability of the particle.

$\hat{V}_E^{nucl}$ describes the effective potential energy of
particle interaction with the pseudoelectric field of the target.

\section{The equations describing the spin evolution of a particle in a storage ring}

 Let us consider particles moving in a storage ring with low
pressure of residual gas ($10^{-10}$ Torr) and without targets
inside the storage ring.
 In this case we can omit the effects
caused by the interaction
$\hat{V}_E^{nucl}$.

 Let us consider a deuteron
moving in a storage ring.
 According to the above analysis spin
behavior of a deuteron can not be described by the
Bargman-Myshel-Telegdy equation.
 The equations for particle spin
motion
in condition when the fields $\vec{E}$ and $\vec{B}$ are
orthogonal to the particle velocity $\vec{v}$
were obtained in  \cite{nastya}-\cite{birefringence}.
They can be written as follows:
\begin{eqnarray}
\left\{
\begin{array}{l}
\frac{d\vec{P}}{dt}=
\frac{e}{mc}\left[\vec{P}\times\left\{\left(a+\frac{1}{\gamma}\right)\vec{B}
-
\left(\frac{g}{2}-\frac{\gamma}{\gamma+1}\right)\vec{\beta}\times\vec{E}\right\}\right]+\\
 +
\frac{d}{\hbar}\left[\vec{P}\times\left({\vec{E}}+\vec{\beta}\times\vec{B}\right)\right]
-\frac{2}{3}\frac{\alpha_{T}E^{2}_{eff}}{\hbar}[\vec{n}_{E}\times\vec{n}_{E}^{\prime}]
-\frac{2}{3}\frac{\beta_{T}B^{2}_{eff}}{\hbar}[\vec{n}_{B}\times\vec{n}_{B}^{\prime}],\\
{} \\
\frac{dP_{ik}}{dt}  =
-\left(\varepsilon_{jkr}P_{ij}\Omega_{r}(d)+\varepsilon_{jir}P_{kj}\Omega_{r}(d)\right)
 - \\
-
\frac{3}{2}\frac{\alpha_{T}E^{2}_{eff}}{\hbar}\left([\vec{n}_{E}\times\vec{P}]_{i}n_{E,\,k}
+n_{E,\,i}[\vec{n}_{E}\times\vec{P}]_{k}\right)-\\
 -
\frac{3}{2}\frac{\beta_{T}B^{2}_{eff}}{\hbar}\left([\vec{n}_{B}\times\vec{P}]_{i}n_{B,\,k}
+n_{B,\,i}[\vec{n}_{B}\times\vec{P}]_{k}\right),
\\
\end{array}
\right. \label{50}
\end{eqnarray}
where $m$ is the mass of the particle, $e$ is its charge,
$\vec{P}$ is the spin polarization vector, $P_ik$ is the spin
polarization tensor, $P_{xx}+P_{yy}+P_{zz}=0$, $\gamma$ is the
Lorentz-factor,
 $\vec{\beta}=\vec{v}/c$,
$\vec{v}$ is the particle velocity,
$a=(g-2)/2$, $g$ is the gyromagnetic ratio, $
\vec{E}$ and
 $\vec{B}$ are the electric and magnetic fields in the point of
 particle location, $\vec{E}_{eff}=(\vec{E}+\vec{\beta} \times
\vec{B})$, $\vec{B}_{eff}=(\vec{B}-\vec{\beta} \times \vec{E})$,
$\vec{n}_{E}=\frac{\vec{E}+\vec{\beta}\times\vec{B}}{|\vec{E}+\vec{\beta}\times\vec{B}|}$,
$\vec{n}_B=\frac{\vec{B}-\vec{\beta} \times
\vec{E}}{|\vec{B}-\vec{\beta} \times \vec{E}|}$,
 $n_{E,\,i}^{\prime}=P_{ik}n_{E,\,k}$,
$n_{Bi}^{\prime}=P_{il}n_{Bl}$, $\Omega_{r}(d)$ are the components
of the vector $\vec{\Omega}(d)$ ($r=1,2,3$ correspond to $x,y,z$,
respectively).
\begin{eqnarray}
\begin{array}{l}
\vec{\Omega}(d)  =  \vec{\Omega} + \vec{\Omega}_{d}, \nonumber \\
\vec{\Omega}  =
\frac{e}{mc}\left\{\left(a+\frac{1}{\gamma}\right)\vec{B} -
\left(\frac{g}{2}-\frac{\gamma}{\gamma+1}\right)\vec{\beta}\times\vec{E}\right\}, \nonumber \\
\vec{\Omega}_{d}  =
\frac{d}{\hbar}\left({\vec{E}}+\vec{\beta}\times\vec{B}\right).
\label{2.24}
\end{array}
\end{eqnarray}
The equations for particle spin motion (\ref{50}) can be rewritten
as follows:
\begin{eqnarray}
\frac{d\vec{P}}{dt}=[\vec{P}\times\vec{\Omega}(d)]+\Omega_T[\vec{n}_E\times\vec{n}_E^{\prime}]+
\Omega_T^{\mu}[\vec{n}_B\times\vec{n}_B^{\prime}], \nonumber\\
{} \nonumber\\
\frac{d\vec{P_{ik}}}{dt}=-(\epsilon_{jkr}P_{ij}\Omega_r(d)+\epsilon_{jir}P_{kj}\Omega_r(d))+
\Omega_T^{\prime}([\vec{n}_E\times\vec{P}]_i
n_{Ek}+n_{Ei}[\vec{n}_E\times\vec{P}]_k) + \nonumber \\
{} \nonumber\\
+\Omega_T^{\prime \mu}([\vec{n}_B\times\vec{P}]_i
n_{Bk}+n_{Bi}[\vec{n}_B\times\vec{P}]_k)\label{BMT} \label{BMT+}
\end{eqnarray}
 where
\begin{eqnarray}
\begin{array}{l}
\Omega_T=-\frac{2}{3} \frac{\alpha_T E_{eff}^2}{\hbar} ,
~~\Omega_T^{\prime}=-\frac{3}{2} \frac{\alpha_T E_{eff}^2}{\hbar},
~~\Omega_T^{\prime}=-\frac{2}{3} \Omega_T,\nonumber \\
{}\nonumber \\
 \Omega_T^{\mu}=-\frac{2}{3} \frac{\beta_T
B_{eff}^2}{\hbar} , ~~\Omega_T^{\prime \mu}=-\frac{3}{2}
\frac{\beta_T B_{eff}^2}{\hbar}, ~~\Omega_T^{\prime
\mu}=-\frac{2}{3} \Omega_T^{\mu}.
\end{array}
\end{eqnarray}

Suppose that the external electric field in the storage ring
$\vec{E}=0$ and a particle moves  {along the circle orbit}.

Let us now consider the equation (\ref{BMT+}) in the coordinate
system that rotates with the frequency of particle velocity
rotation.
 {In such a system spin rotates with respect to the momentum
with the frequency determined by $(g-2)$.}%
The coordinate system and vectors $\vec{v},\vec{E}, \vec{B}$ as
shown in figure and denote the axes by $x,y,z$ (or $1,2,3$,
respectively).
\begin{figure}[!h]
\begin{center}
\includegraphics[width=8cm,keepaspectratio]{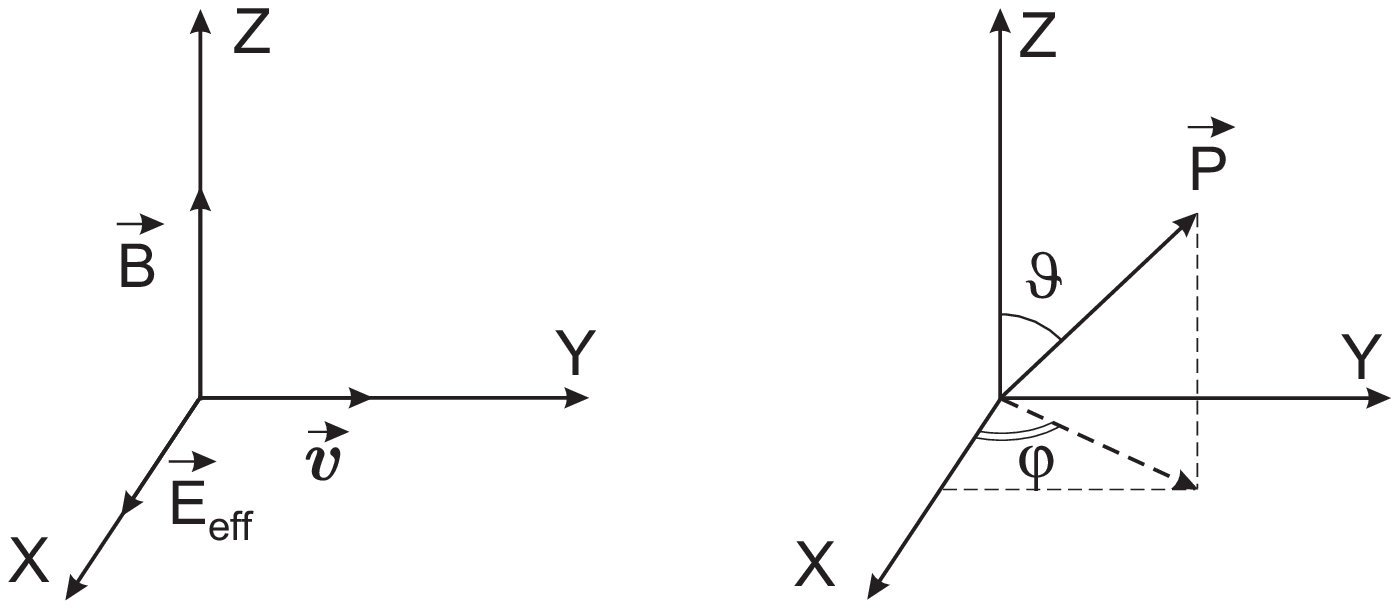}
\caption{} \label{coordinate}
\end{center}
\end{figure}

Therefore, the components of the vectors are:
\begin{eqnarray}
\begin{array}{l}
\vec{P}=\left(P_1,P_2,P_3\right), \\
  {}  \\
 \vec{n}_E=\left(1,0,0\right), n_{Ei}^{\prime}=P_{il}n_{El}=P_{i1}
   \\
  {} \\
 {[} \vec{n}_E\times\vec{n}_{E}^{\prime}{]}_1=0,~{[}\vec{n}_E\times\vec{n}_{E}^{\prime}{]}_2=-P_{31},~
 {[}\vec{n}_E\times\vec{n}_{E}^{\prime}{]}_3=P_2,
 \\
{}  \\
 {[}\vec{P}\times\vec{\Omega} {]}_1=\Omega P_2,~
 {[} \vec{P}\times\vec{\Omega}{]}_2=-\Omega P_1,~
 {[} \vec{P}\times\vec{\Omega}{]}_3=P_2, \\
{}\\
 {[} \vec{n}_E\times\vec{P}{]}_1=0,~
 {[} \vec{n}_E\times\vec{P}{]}_2=-P_3,~
 {[} \vec{n}_E\times\vec{P}{]}_3=P_2,
\end{array}
\label{E}
\end{eqnarray}
\begin{eqnarray}
\begin{array}{l}
\vec{\Omega}=\frac{ea}{mc}\vec{B}=\left(0,0,\Omega\right),
 \\
  {} \\
 \vec{n}_B=\left(0,0,1\right), n_{Bi}^{\prime}=P_{il}n_{Bl}=P_{i3}
 \\
  {} \\
 {[}\vec{n}_B\times\vec{n}_{B}^{\prime}{]}_1=-P_{23},~{[}\vec{n}_B\times\vec{n}_{B}^{\prime}{]}_2=-P_{13},~
 {[} \vec{n}_B\times\vec{n}_{B}^{\prime}{]}_3=0,
\\
{} \\
 {[} \vec{n}_B\times\vec{P}{]}_1=-P_2,~
 {[} \vec{n}_B\times\vec{P}{]}_2=P_1,~
 {[} \vec{n}_B\times\vec{P}{]}_3=0.
 \label{B}
\end{array}
\end{eqnarray}

Substituting (\ref{E},\ref{B}) to the system (\ref{50}) we obtain:
\begin{eqnarray}
\begin{array}{l}
\frac{dP_1}{dt}=\Omega P_2-\Omega_T^{\mu} P_{23},\\
{}  \\
 \frac{dP_2}{dt}=-\Omega P_1+(\Omega_T^{\mu}-\Omega_T)P_{13}+\omega_d P_3,\\
 {}  \\
 \frac{dP_3}{dt}=\Omega_T P_{12}-\omega_d P_2,
 \end{array}
\label{Pi}
\end{eqnarray}
where $\omega_d=\frac{dE_1^{eff}}{\hbar}$.
\begin{eqnarray}
\begin{array}{l}
\frac{dP_{11}}{dt} = 2 \Omega P_{12}+2 \omega_d P_{23},  \\
 {}  \\
\frac{dP_{22}}{dt} = -2 \Omega P_{12}, \\
 {}  \\
\frac{dP_{33}}{dt} =-2 \omega_d P_{23}, \\
 \end{array}
\label{sys1}
\end{eqnarray}
\begin{eqnarray}
\begin{array}{l}
\frac{dP_{12}}{dt} = -\Omega \left(
P_{11}-P_{22}\right)-\Omega_T^{\prime} P_3+\omega_d P_{13},
 \\
 {}  \\
\frac{dP_{13}}{dt} = \Omega P_{23}+{\Omega}_T^{\prime}P_2-{\Omega}_T^{\prime \mu}P_2 -\omega_d P_{12},  \\
{}  \\
\frac{dP_{23}}{dt} = -\Omega P_{13}+{\Omega}_T^{\prime \mu}P_1
-\omega_d (P_{22}-P_{33}).
 \end{array}
\label{p12def}
\end{eqnarray}
Remember that $P_{11}+P_{22}+P_{33}=0$ and $P_{ik}=P_{ki}$.

\section{Contribution from the EDM and tensor polarizabilities to
deuteron spin oscillation}

Let us consider the system (\ref{Pi}-\ref{p12def}) more
attentively.

Suppose that deuteron has neither EDM no tensor electric
polarizability: in this case the system (\ref{Pi}-\ref{p12def})
can be expressed:
\begin{eqnarray}
\begin{array}{l}
\frac{dP_1}{dt}=\Omega P_2,\\
{}  \\
 \frac{dP_2}{dt}=-\Omega P_1,\\
 {}  \\
 \frac{dP_3}{dt}=0,
 \end{array}
\label{Pi0}
\end{eqnarray}
\begin{eqnarray}
\begin{array}{l}
\frac{dP_{11}}{dt} = 2 \Omega P_{12},  \\
 {}  \\
\frac{dP_{22}}{dt} = -2 \Omega P_{12}, \\
 {}  \\
\frac{dP_{33}}{dt} =0, \\
 \end{array}
\label{sys10}
\end{eqnarray}
\begin{eqnarray}
\begin{array}{l}
\frac{dP_{12}}{dt} = -\Omega \left( P_{11}-P_{22}\right),
 \\
 {}  \\
\frac{dP_{13}}{dt} = \Omega P_{23},  \\
{}  \\
\frac{dP_{23}}{dt} = -\Omega P_{13}.
 \end{array}
\label{p12def0}
\end{eqnarray}
This is the conventional system of BMT equations that describes
particle spin rotation with the frequency equal to $(g-2)$
precession frequency ${\Omega}=\frac{ea}{mc}{B}$.
The component $P_3$ of vector polarization in this conditions is
equal to constant $(\frac{dP_{3}}{dt} =0)$ along with the
component $P_{33}$ of tensor polarization  $(\frac{dP_{33}}{dt}
=0)$.

Suppose the deuteron EDM differs from zero. Then the above system
of equations converts to:
\begin{eqnarray}
\begin{array}{l}
\frac{dP_1}{dt}=\Omega P_2,\\
{}  \\
 \frac{dP_2}{dt}=-\Omega P_1+\omega_d P_3,\\
 {}  \\
 \frac{dP_3}{dt}=-\omega_d P_2,
 \end{array}
\label{PiE}
\end{eqnarray}
\begin{eqnarray}
\begin{array}{l}
\frac{dP_{11}}{dt} = 2 \Omega P_{12}+2 \omega_d P_{23},  \\
 {}  \\
\frac{dP_{22}}{dt} = -2 \Omega P_{12}, \\
 {}  \\
\frac{dP_{33}}{dt} =-2 \omega_d P_{23}, \\
 \end{array}
\label{sys1E}
\end{eqnarray}
\begin{eqnarray}
\begin{array}{l}
\frac{dP_{12}}{dt} = -\Omega \left( P_{11}-P_{22}\right)+\omega_d
P_{13},
 \\
 {}  \\
\frac{dP_{13}}{dt} = \Omega P_{23} -\omega_d P_{12},  \\
{}  \\
\frac{dP_{23}}{dt} = -\Omega P_{13} -\omega_d (P_{22}-P_{33}).
 \end{array}
\label{p12defE}
\end{eqnarray}
From  (\ref{PiE}-\ref{p12defE}) it follows that presence of the
nonzero EDM makes the vertical component $P_3$ of vector
polarization oscillating with the frequency of $(g-2)$ precession
$\Omega$.

According to the idea \cite{orlov} these oscillations can be
eliminated if the deuteron velocity is modulated with the
frequency $\Omega$:
\begin{equation}
v=v_0+\delta v \sin{(\Omega t + \varphi_f)} \label{v}
\end{equation}
here $\varphi_f$ is the phase of forced oscillations of the
velocity.

As  $E_{eff}$ depends on $\vec{\beta}=\vec{v}/c$ it also appears
modulated:
\begin{equation}
E_{eff}=E_{eff}^0+\delta E_{eff} \sin{(\Omega t + \varphi_f)}
\end{equation}
Therefore $\omega_d=\frac{d E_{eff}}{\hbar}$ is also modulated with the same frequency.
This makes the product $\omega_d P_2 \sim \sin^2 (\Omega t+ \varphi_f)$.
Therefore averaging this value over the period of $(g-2)$
precession gives the result time-independent (i.e.
$\frac{dP_3}{dt}=const$) and $P_3 (t)=P_3 (0)+const \cdot t$.
For better measurement conditions it is important to make $P_3
(0)=0$.
 {This is the reason to chose particle spin to be in the
horizontal plane.}

All the above reasoning makes $\frac{dP_{33}}{dt} \sim const$,
too.
Therefore, $P_{33}$ also linearly grows with time $P_{33}
(t)=P_{33} (0)+const \cdot t$.
However, if the spin lays in the horizontal plane $P_{33}(0) \ne
0$.

It is important to note (see below the section \ref{sec:4.2}) that
if the spin orientation corresponds to $\cos^2 \vartheta =
\frac{1}{3}$ ($\cos \vartheta =\sqrt{\frac{1}{3}}, ~ \sin
\vartheta = \sqrt{\frac{2}{3}}$) then the component $P_{33} (0) =
0$, while $P_3(0) \ne 0$.

Therefore taking $\vartheta$ corresponding to $\cos \vartheta=
\sqrt{\frac{1}{3}}$ we can use the component $P_{33}$ for EDM
measurements, too.


Let us consider the contribution from the electric and magnetic
tensor polarizabilities. Then instead the system
(\ref{PiE}-\ref{p12defE}) we should consider the system
(\ref{Pi}-\ref{p12def})

\begin{eqnarray}
\begin{array}{l}
\frac{dP_1}{dt}=\Omega P_2-\Omega_T^{\mu} P_{23},\\
{}  \\
 \frac{dP_2}{dt}=-\Omega P_1+(\Omega_T^{\mu}-\Omega_T)P_{13}+\omega_d P_3,\\
 {}  \\
 \frac{dP_3}{dt}=\Omega_T P_{12}-\omega_d P_2,
 \end{array}
\label{Pi2}
\end{eqnarray}
\begin{eqnarray}
\begin{array}{l}
\frac{dP_{11}}{dt} = 2 \Omega P_{12}+2 \omega_d P_{23},  \\
 {}  \\
\frac{dP_{22}}{dt} = -2 \Omega P_{12}, \\
 {}  \\
\frac{dP_{33}}{dt} =-2 \omega_d P_{23}, \\
 \end{array}
\label{sys2}
\end{eqnarray}
\begin{eqnarray}
\begin{array}{l}
\frac{dP_{12}}{dt} = -\Omega \left(
P_{11}-P_{22}\right)-\Omega_T^{\prime} P_3+\omega_d P_{13},
 \\
 {}  \\
\frac{dP_{13}}{dt} = \Omega P_{23}+{\Omega}_T^{\prime}P_2-{\Omega}_T^{\prime \mu}P_2 -\omega_d P_{12},  \\
{}  \\
\frac{dP_{23}}{dt} = -\Omega P_{13}+{\Omega}_T^{\prime \mu}P_1
-\omega_d (P_{22}-P_{33}).
 \end{array}
\label{p12def2}
\end{eqnarray}
Some interesting implications follow from
(\ref{Pi2}-\ref{p12def2}).
 As it was already mentioned above in the experiments for EDM
search it is planned to measure growth of the vertical component
of the polarization vector $P_3$.

According to (\ref{Pi2}) time dependence of the vertical component
of the vector polarization $P_3$ is described by the equation
\begin{equation}
\frac{dP_3}{dt}=\Omega_T P_{12}-\omega_d P_2
\end{equation}
As it can be seen the time dependence of $P_3$ is determined by
both the EDM and tensor polarizability of deuteron.
It is interesting that the derivative of the tensor polarization
component $P_{33}$ does not contain contributions from tensor
electric polarizability and is proportional to the EDM only:
\begin{equation}
\frac{dP_{33}}{dt} =-2 \omega_d P_{23}
\end{equation}
Therefore, it is  important to measure the component $P_{33}$,
too.
According to the above spin orientation for this case is
determined by the condition  $\cos^2 \vartheta = \frac{1}{3}$.


\subsection{Contribution from the tensor magnetic polarizability to deuteron spin oscillation}

Contributions to spin rotation and oscillations from EDM and
polarizabilities are small. Therefore, they, being analyzed, could
be considered as perturbations to the full system
(\ref{Pi2}-\ref{p12def2}) and the role of each could be studied
separately.

The system of equations considering contribution from the tensor
magnetic polarizability $\beta_T$ is as follows:
\begin{eqnarray}
\begin{array}{l}
\frac{dP_1}{dt}=\Omega P_2-\Omega_T^{\mu} P_{23},\\
{}  \\
 \frac{dP_2}{dt}=-\Omega P_1+\Omega_T^{\mu} P_{13},\\
 {}  \\
 \frac{dP_{13}}{dt} = \Omega P_{23}-{\Omega}_T^{\prime \mu}P_2,  \\
{}  \\
\frac{dP_{23}}{dt} = -\Omega P_{13}+{\Omega}_T^{\prime \mu}P_1
\end{array} \label{sys2}
\end{eqnarray}

Introducing new variables $P_{+}=P_1+iP_2$ and
$G_{+}=P_{13}+iP_{23}$ and recomposing equations (\ref{sys2}) to
determine $P_{+}$ and $G_{+}$ we obtain:
\begin{eqnarray}
\begin{array}{l}
\frac{dP_+}{dt}=-i\Omega P_+ + i\Omega_T^{\mu} G_+,\\
{} \nonumber \\
 \frac{dG_+}{dt}=-i\Omega G_+ + i\Omega_T^{\prime \mu} P_+,\\
\end{array}
\label{sys22}
\end{eqnarray}

or

\begin{eqnarray}
\begin{array}{l}
i\frac{dP_+}{dt}=\Omega P_+ - \Omega_T^{\mu} G_+,\\
{} \nonumber \\
i \frac{dG_+}{dt}=\Omega G_+ - \Omega_T^{\prime \mu} P_+,\\
\end{array}
\label{sys3}
\end{eqnarray}

Let us search $P_+,G_+ \sim e^{i\omega t}$ then (\ref{sys3})
transforms as follows:

\begin{eqnarray}
\begin{array}{l}
\omega \tilde{P}_+=\Omega \tilde{P}_+ - \Omega_T^{\mu} \tilde{G}_+,\\
{} \nonumber \\
\omega \tilde{G}_+=\Omega \tilde{G}_+ - \Omega_T^{\prime \mu}
\tilde{P}_+.
\end{array}
\label{sys33}
\end{eqnarray}
The solution of this system can be easily find:
\begin{eqnarray}
\begin{array}{l}
(\omega -\Omega)^2- \Omega_T^{\mu} \Omega_T^{\prime \mu}=0\\
\end{array}
\label{sys34}
\end{eqnarray}
 that finally gives
 \begin{equation}
\omega_{1,2}=\Omega \pm \sqrt{\Omega_T^{\mu} \Omega_T^{\prime
\mu}}
\end{equation}
Rewriting the solution
 \begin{equation}
P_+(t)=c_1 e^{-i\omega_1 t}+c_2 e^{-i\omega_2 t}= |c_1|
e^{-i(\omega_1 t-\delta_1)}+|c_2| e^{-i(\omega_2 t-\delta_2)}
\label{P+}
\end{equation}
Therefore,
 \begin{equation}
P_1(t)=|c_1| \cos (\omega_1 t - \delta_1)+|c_2| \cos (\omega_2 t -
\delta_2) \label{reP+}
\end{equation}
 \begin{equation}
P_2(t)=-|c_1| \sin (\omega_1 t - \delta_1)-|c_2| \sin (\omega_2 t
- \delta_2) \label{imP+}
\end{equation}


According to (\ref{reP+},\ref{imP+})the nonzero deuteron tensor
magnetic polarizability makes the spin rotating with two
frequencies $\omega_1$ and
 $\omega_2$ instead of $\Omega$ and, therefore, experiences beating with the frequency
$\Delta \omega=\omega_1-\omega_2=2 \sqrt{\Omega_T^{\mu}
\Omega_T^{\prime \mu}}=\frac{\beta_T B_{eff}^2}{\hbar} $.

Let us recall now that EDM interaction with the electric field
makes the deuteron spin rotating around the direction of this
field and leads to appearance of $P_3$ component proportional to
$P_2(t)$
 \begin{equation}
 \frac{dP_3}{dt} \sim - \omega_d P_2
\end{equation}

Therefore,

 \begin{equation}
 \frac{dP_3}{dt}=\omega_d \left( |c_1| \sin (\omega_1 t - \delta_1)+|c_2| \sin (\omega_2
t - \delta_2) \right)
\end{equation}

According to the idea \cite{orlov} to measure the EDM the particle
velocity ($v=v_0+\delta v \sin{(\Omega_f t + \varphi_f)}$) should
be modulated with the frequency $\Omega_f$ close to the frequency
$\Omega$ of $(g-2)$ precession.

If the magnetic polarizability is equal to zero, then $\omega_1 =
\omega_2=\Omega$ and spin rotates in the horizontal plane with the
frequency $\Omega$. In this case velocity modulation with the same
frequency $\Omega_f=\Omega$ gives
 \begin{equation}
 \frac{dP_3}{dt} \sim \sin^2 (\Omega t)
\end{equation}
and the vertical component $P_3$ linearly grows with time.

However, $\omega_1 \ne \omega_2$ and velocity modulation, for
example, with the frequency $\Omega=\omega_1$ provides for slow
spin oscillation with the frequency $\omega_1 - \omega_2$ instead
of linear growth.

According to the evaluation \cite{polarizability} the tensor
magnetic polarizability $\beta_T \sim 2 \cdot 10^{-40}$, therefore
the beating frequency $\Delta \omega \sim 10^{-5}$ in the field $B
\sim 10^4$ gauss.

Measurement of the frequency of this beating makes possible to
measure the tensor magnetic polarizability of the deuteron
(nuclei).

Thus, due to the presence of tensor magnetic polarizability the
the horizontal component of spin rotates around $\vec{B}$ with two
frequencies $\omega_1,~\omega_2$ instead of expected rotation with
the frequency $\Omega$.

This is the reason for the component $P_3$ caused by the EDM to
experience the similar oscillations.
Therefore, particle velocity modulation with the frequency
$\Omega$  provides for eliminating oscillation with $\Omega$
frequency, but $P_3$ oscillations with the frequency $\Delta
\omega$ rest (similarly $P_{33}$).
Study of these oscillations is necessary because they can distort the EDM measurements.

\subsection{Contribution from the tensor electric polarizability to deuteron spin oscillation}
\label{sec:4.2}

Let us consider now contribution caused by the tensor electric
polarizability.
From the system (\ref{sys1}) it follows

\begin{eqnarray}
\begin{array}{l}
\frac{d(P_{11}-P_{22})}{dt} = 4\Omega P_{12},
\\
 {}  \\
\frac{d^2P_{12}}{dt^2} = -\Omega
\frac{d(P_{11}-P_{22})}{dt}-{\Omega}_T^{\prime} \frac{dP_3}{dt}=
-(4\Omega^2+\Omega_{T}{\Omega}_T^{\prime})P_{12}.  \\
 \end{array}
\label{p11-p22}
\end{eqnarray}

Thus we have the equation
\begin{equation}
\frac{d^2P_{12}}{dt^2}+ \omega_{12}^2 P_{12}=0
\end{equation}
where $\omega_{12}=\sqrt{4\Omega^2+\Omega_{T}{\Omega}_T^{\prime}}
\approx 2 \Omega$, because $\Omega_{T}{\Omega}_T^{\prime} \ll
\Omega^2$.

The solution for this equation can be found in the form:
\begin{equation}
P_{12}=c_1 \cos{\omega_{12} t}+c_2 \sin{\omega_{12} t} \label{P12}
\end{equation}

Let us find coefficients $c_1$ and $c_2$: when $t=0$ the equation
(\ref{P12}) gives $c_1=P_{12}(0)$. The coefficient $c_2$ can be
found from
\begin{equation}
\frac{d(P_{12})}{dt}(t \rightarrow 0)=\omega_{12} c_2,
\end{equation}
therefore
\begin{equation}
c_2= \frac{1}{\omega_{12}} \frac{d(P_{12})}{dt}(t \rightarrow 0) ,
\end{equation}
From the equation (\ref{p12def})
\begin{equation}
\frac{dP_{12}}{dt} (t \rightarrow 0)= -\Omega \left( P_{11}(t
\rightarrow 0)-P_{22}(t \rightarrow 0)\right), \label{c2}
\end{equation}
that
\begin{equation}
c_2=-\frac{P_{11}-P_{22}}{2}, \label{c2a}
\end{equation}
and
\begin{equation}
P_{12}=P_{12}(0) \cos{\omega_{12} t}-\frac{P_{11}-P_{22}}{2}
\sin{\omega_{12} t} \label{P12final}
\end{equation}

As a result we can write the following equation for the vertical
component of the spin $P_3$:
\begin{equation}
\frac{dP_{3}}{dt}=\Omega_T P_{12}(t)= \Omega_T [ P_{12}(0)\cos{2
\Omega t}-\frac{P_{11}(0)-P_{22}(0)}{2} \sin{2\Omega t} ]
\label{P12final1}
\end{equation}

As it can be seen the vertical component of the spin oscillates
with the frequency $2 \Omega$.

But it should be reminded that according to the equations
(\ref{50}) interaction of the EDM with an electric field causes
oscillations of the vertical component of the spin with the
frequency $\Omega$. According to the idea \cite{orlov} these
oscillations can be eliminated if the deuteron velocity is
modulated with the frequency $\Omega_f$ that should be taken as close to the frequency $\Omega$ as possible:
\begin{equation}
v=v_0+\delta v \sin{(\Omega_f t + \varphi_f)} \label{v}
\end{equation}
 As a result $E_{eff}$ depends on $\vec{\beta}=\vec{v}/c$ it also appears
modulated:
\begin{equation}
E_{eff}=E_{eff}^0+\delta E_{eff} \sin{(\Omega_f t + \varphi_f)}.
\end{equation}

The equation describing contribution  from the tensor electric
polarizability and EDM to $P_3$ looks like (\ref{Pi})
\begin{equation}
\frac{dP_3}{dt}=\Omega_T P_{12} - \omega_d P_2.
\end{equation}
As $P_2$ oscillates with the frequency $\Omega$,
 {then the product $\omega_d P_2$ contains the
non-oscillating terms} and contribution to $P_3$ caused by EDM
linearly grows with time, if $\Omega_f=\Omega$.
If $\Omega_f \ne \Omega$ contribution to $P_3$ caused by EDM
slowly oscillates with the frequency $\Omega_f - \Omega$.

It is important that modulation of the velocity $v=v_0+\delta v
\sin{(\Omega_f t + \varphi_f)}$ results in oscillation of
$E_{eff}^2$ also oscillates with time and appears proportional to
$\sin^2 {(\Omega_f t + \varphi_f)}$.
 As a result the contribution to $P_3$ caused by the tensor
 electric polarizability can be expressed as:
\begin{equation}
\frac{dP_3}{dt} \sim \Delta \Omega_T \sin^2 {(\Omega_f t +
\varphi_f)}[ P_{12}(0)\cos{2 \Omega
t}-\frac{P_{11}(0)-P_{22}(0)}{2} \sin{2\Omega t} ]
\end{equation}
 i.e.
\begin{equation}
\frac{dP_3}{dt} \sim - \frac{1}{2} \Delta \Omega_T \cos {(2
\Omega_f t+ 2\varphi_f)}[ P_{12}(0)\cos{2 \Omega
t}-\frac{P_{11}(0)-P_{22}(0)}{2} \sin{2\Omega t} ] \label{28}
\end{equation}

According to (\ref{28}) for a partially polarized deuteron beam
the derivative $\frac{dP_3}{dt}$
depends on the deuteron polarization components $P_{12}$ and
$\frac{P_{11}(0)-P_{22}}{2}$


For simplicity let us consider a deuteron beam in pure
polarization state.
In this case the components $P_{12}$ and
$\frac{P_{11}(0)-P_{22}}{2}$ can be written using the explicit
expression for the spin wavefunctions.
Suppose $\vec{n}(\vartheta, \varphi)$ is the unit vector directed
along the deuteron spin  ($\vartheta$ and $\varphi$ are the polar
and azimuth angles (see Fig.\ref{coordinate})).
  So the spin wavefunction that
describes the deuteron spin state with the magnetic quantum number
$m=1$
  can be expressed as follows (in the Cartesian
 basis):
\begin{eqnarray}
\chi_{1}(\vartheta, \varphi)=\left(
\begin{array}{c}
a_x\\
a_y\\
a_z
\end{array}
\right)=-\frac{1}{\sqrt{2}} \left(
\begin{array}{c}
\cos \vartheta \cos \varphi - i \sin \varphi\\
\cos \vartheta \sin \varphi + i \cos \varphi\\
- \sin \vartheta
\end{array}
\right)
\end{eqnarray}

Polarization vector can be written as
\begin{equation}
\vec{P}=\langle \hat{S} \rangle=\chi_1^+ \hat{S} \chi_1 = i
[\vec{a} \times \vec{a}^{*}]
\end{equation}
and components of polarization tensor
\begin{equation}
\langle {P}_{ik}\rangle=\chi_1^+ \hat{Q}_{ik} \chi_1= -\frac{3}{2}
\{ a_i a_k^* + a_k a_i^* - \frac{2}{3}\},
\end{equation}
where $\hat{Q}_{ik}$ is the spin tensor of rank two.
Therefore,
\begin{eqnarray}
P_{12}=\frac{3}{4} \sin 2 \varphi \sin^2 \vartheta, \\
\frac{P_{11}-P_{22}}{2}=\frac{3}{4} \cos 2 \varphi \sin^2
\vartheta, \\
P_{33}=-\frac{3}{2} \left( \sin^2 \vartheta - \frac{2}{3}\right).
\end{eqnarray}

Using (49,53,54) one can obtain:
\begin{eqnarray}
\frac{dP_3}{dt} \sim - \frac{3}{8} \Delta \Omega_T \sin^2
\vartheta \cos (2 \Omega_f t + 2 \varphi_f) \times [\sin 2 \varphi
\cos 2 \Omega t - \cos 2 \varphi \sin 2 \Omega t] \label{101}
\end{eqnarray}
From (\ref{101}) it follows that $\frac{dP_3}{dt}$ slowly
oscillates with the frequency $(\Omega_f - \Omega)$

In the ideal case, when $\Omega_f = \Omega$ (as it is proposed in
\cite{orlov} for EDM measurement) (\ref{101}) converts to
\begin{eqnarray}
\frac{dP_3}{dt} = - \frac{3}{8} \Delta \Omega_T \sin^2 \vartheta
\cos (2 \Omega t + 2 \varphi_f) \sin ( 2 \Omega t -  2 \varphi)
\label{102}
\end{eqnarray}
In the general case, when the phases $\varphi_f$ and $\varphi$ are
arbitrary, (\ref{102}) contains terms that do not depend on time
and, therefore, $P_3$ linearly grows with time like the signal
from the EDM does.

It is interesting that making $\varphi_f=-\varphi$ gives
\begin{eqnarray}
\frac{dP_3}{dt}  \sim \cos (2 \Omega t - 2 \varphi) \sin ( 2
\Omega t -  2 \varphi)=\frac{1}{2} \sin (4 \Omega t - 4 \varphi)
\label{103}
\end{eqnarray}
 {that makes this contribution to $P_3$ quickly oscillating
and depressed}.
But even in this ideal case  it rests the contribution caused by
the tensor magnetic polarizability (\ref{VB}), in real situation
$\Omega \ne \Omega_f$, though.

Measurement of these contribution provides to measure the tensor
electric polarizability.

According to the evaluations \cite{polarizability} $\alpha_T \sim
10^{-40}$ cm$^3$  ,
therefore for the field $E_{eff}=B \sim 10^4$ gauss the frequency $ \Omega_T \sim 10^{-5}$ sec$^{-1}$.
When
considering modulation we should estimate $\Delta \Omega_T \sim
\Omega_T {(\frac{\delta}{v_0})}^2$, then  suppose
${(\frac{\delta}{v_0})}^2 \sim 10^{-2} - 10^{-3}$ we obtain
$\Delta \Omega_T \sim 10^{-7} - 10^{-8}$ sec$^{-1}$,
 {that exceeds the magnitude of $\omega_d$ for the deuteron
EDM $d=10^{-29}~e \cdot cm$}.

\section{Conclusion}

In the present paper influence of tensor electric and
 magnetic polarizabilities on spin evolution in the resonance deuteron EDM
 experiment  is considered in details.

 It is shown that besides EDM the electric and magnetic polarizabilities also contribute
 to the vertical spin component $P_3$. Moreover, the electric polarizability contributes
 to the $P_3$ component even when the deuteron EDM is supposed to be zero and thereby
 the electric polarizability can imitate the EDM contribution.
 It is shown that unlike the vertical component of the spin $P_3$
 the component $P_{33}$ of polarization tensor does not contain
 contribution from the electric polarizability, whereas contribution from the magnetic
 polarizability reveals only when the deuteron EDM differs from zero.

Moreover, it is also shown that when the angle $\vartheta$ between the spin
direction and the vertical axis meets the condition $\sin
\vartheta=\sqrt{\frac{2}{3}}$ ($\cos
\vartheta=\sqrt{\frac{1}{3}}$), the initial value of $P_{33}$
appears $P_{33}(0)=0$.
As a result, EDM contribution to the measured signal linearly
growth in time starting from zero that is important for
measurements.

Therefore, measurement of the $P_{33}$ component of deuteron
tensor polarization seems to be of particular interest, especially
because  the nonzero component $P_{33}$ appearance on its own
indicates the EDM presence (in contrast to the $P_3$ component,
which appearance can be aroused by the tensor electric
polarizability, rather than EDM).

\end{document}